# SUPRESSION OF FLASHOVER ON RADIO-FREQUENCY WINDOW OF LINEAR ACCELERATOR AT THE TEMPERATURE OF 4.2 K

Xinmeng Liu

**ABSTRACT**

During the operation of the China ADS Front End Superconducting Demo Linac of the China initiative Accelerator Driven System (CiADS), it was found that there was flashover happened on the surface of vacuum ceramic window in the pick-up coupler of superconducting cavity, which is uncommon under low temperature conditions. That seriously affects on the stable operation of the superconducting linear accelerator. By comparing the flashover trajectory on the electrical aging windows from the accelerator, the simulation results of the field emission electrons trajectory in the superconducting cavity, as well as online experiments, we found that the flashover on the window surface was caused by the bombardment and accumulation of field emission electrons from the SRF cavity. Our results suggest that, unlike most vacuum window flashovers, of which characteristic and causation can be explained by the mechanism of the secondary electron emission avalanche (SEEA) model. Some phenomena of the flashover in the linac have been initially explained under the guidance of the mechanism discussed in the PER flashover model. According to the conclusions, a new design modification to the pick-up coupler was made and applied to the linac. The Pt (the power coming out the transmitted power coupler) signal flicker problem has not been monitored after the online application of our newly designed antenna. The research in this article solved the flashover problem and improved the performance of stable operation of the accelerator.

## INTRODUCTION

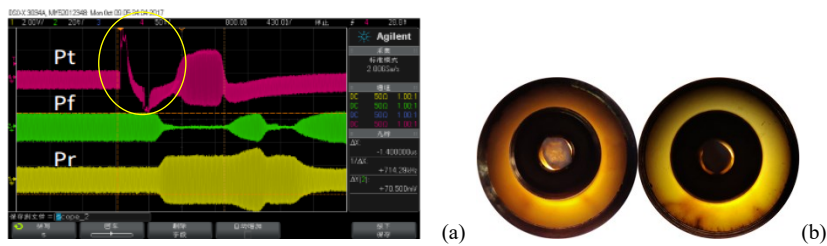

Figure1: The signal on oscilloscope of LLRF system and discharge channel on window.

The China Initiative Accelerator Driven Subcritical System (CiADS) is a strategic plan implemented in China for nuclear waste transmutation [1.2]. The system aims to solve the problem of nuclear waste and realize the recycling of nuclear energy for nuclear power plants. A superconducting linear accelerator that can accelerator a 25MeV, 10 mA proton beam as a demo research facility for the system has been established in order to overcome the challenges in high stability operation technology for sustaining transmutation. As the key components of the linac, the superconducting section, working at 4.2 Kelvin, starts from the HWR010 (half-wavelength β=0.1 microwave resonator) type of Superconducting Radio Frequency (SRF) acceleration cavities, and its running performance determines the capability of the entire accelerator.

In linac, the pick-up coupler is the only reliable way to transmit the cavity RF field information ($P_t$ signal) to the control system. The interference of the deviation signal to the control system will cause the loss of beam, the operation stop and even the damage of the accelerator. However, during the operation of the linear accelerator, an abnormal signal sent from the pickup coupler of the HWR010 SRF cavity is found in the RF control system.

The time-domain waveform diagram of the signal captured by an oscilloscope are shown in Fig. 1(a). For $P_t$ signal the the occurrence of severe fluctuations blocks the previous electromagnetic field which indicate that a discharge has occurred, since a low-resistance breakdown path must be developed when discharge occurs. It can be formed within a few nanoseconds across several centimeters [3]. At the meantime, the forward power ($P_f$) signal and reverse power ($P_r$) signal are remain normal, which indicates that the electromagnetic field in the cavity is not disturbed. The subsequent changes of the two signals are due to the mismatch of the cavity caused by the control system affected by the abnormal $P_t$ signal. Thus, it is reasonable that there is discharge located at the $P_t$ coupler.

In addition, to maintain superconducting state for long time operation, after buffer chemistry polishing, the HWR010 type[4] SRF cavity achieved the surface roughness of the order of 0.5um. Then, the cavities were assembled in a class 100 clean room and within a vacuum environment maitain $1\times10^{-7}$Pa. There is no vulnerable area for breakdown in the vacuum space of coaxial structure SRF cavity, except the vacuum RF window in pick-up coupler provides material interface. The required voltage is considerably less than would be needed to cause bulk breakdown of the insulator or flashover of the vacuum gap if the insulator was removed, since its bulk voltage hold off capability normally is better than that of a similar-sized vacuum gap [5]. Therefore, the discharge should occur on the surface of ceramic window of the pick-up coupler.

To confirm our assumption, the cavities removal check was performed, the results are shown in Fig. 1(b). As expected, traces of discharge on the ceramic windows and channels on the surface of the windows are observed. Moreover, some reticulated channels on the dark area of the window are found under the observation of an optical microscope, which strongly indicated that the flashover happened on the PT coupler windows[6].

**SET UP：**

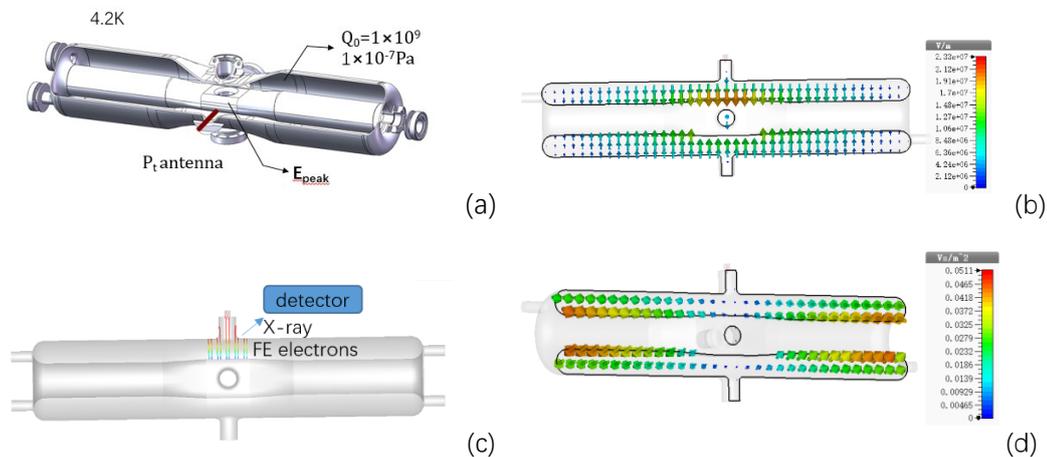

(a)  (b)

(c)  (d)

Fgure2: In order to maintain superconducting state and extremely low energy loss ((quality factor $Q_0=1\times10^9$), HWR010 type cavities work at the temperature of 4.2 K with the vaccum at $1\times10^{-7}$ Pa of the internal environment. The HWR010 SRF cavity working within 162.5MHz TEM mode RF field, the electric field is strongest in the middle and weakest at the both ends which is opposite with the magnetic field. The peak electric field on the surface can reach more than 20MV/m. A photon radiation detector monitors the intensity of the field emission effect to prevent the cavity from the influence of field emission electrons. The coaxial structure pickup coupler with a transmitted power probe extracts a little power (coupling degree β=1/1000) to the RF control system.

The events of the Pt signal spikes were monitored under different field strength applying to cavity, the control system automatically records the moment of each event and the oscilloscope captures the $P_t$ signal spiking waveform. The information about the interval between two events, the peak-to-peak value of the signal, and the duration of the flash were obtained in these records.

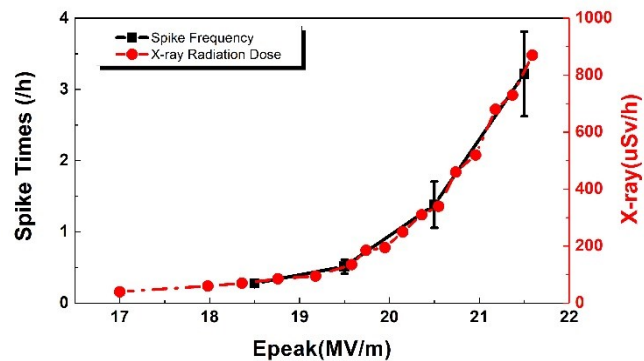

Fgure3: The peak electric field Epk versus the spike Frequency and the X-ray radiation dose. For low Epk, there is no Spike detected. When the Epk is 18.5MV/m, the dose value is 63.55uSv/h, the corresponding average interval between events occurred is 217 minutes. As Epk increases, spike frequency increases, either. In the figure, the flash time is expressed as the frequency of occurrence per unit time, that is, the reciprocal unit of the interval time is times/minute.

The experimental results are given in figure 3. Here, the peak electric field $E_{peak}$ in the cavity is used to characterize the electromagnetic field strength in the cavity. The photon dose under each experimental gradient shows that the magnitude of the current increases with the gradient. Beside, it disappears when the electric field is small to a certain value. This is consistent with the FN theory's [7] description of the properties of field emission currents and the nature of the field emission requirements that determines the field threshold.[8] The experimental results show that flashing occurs more frequently as $E_{peak}$ increases. On the other hand, the secondary electron multiplication does not have the above characteristics and usually occurs in low field conditions. The experimental results show that flashing occurs more frequently with $E_{peak}$ increases. Thus, the experimental evidence suggests that the occurrence of flashover is inseparable from the contribution of field emission.

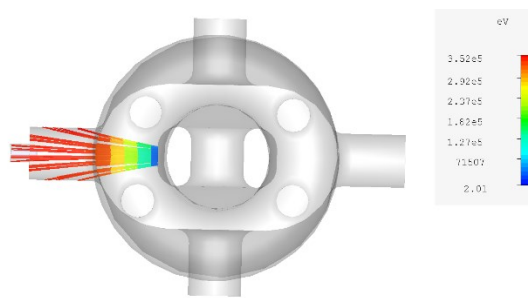

Fgure4: The field emission electrons reach the ceramic window at the action of RF field.

In order to understand the trajectory and final position of the FE electrons in the electromagnetic field of the superconducting cavity. Simulation of the movement of the excitation FE under the action of the electromagnetic field of the cavity at $E_{peak}$ 19MV/m was performed, the results is shown in Figure 4. In the work area, the electric field is strong, while the magnetic field extremely weak and has almost no deflection effect on electrons. Thus, the FE electrons move straight along the electric field lines before they bombard the ceramic window surface and cavity wall with an energy of more than 0.3MeV. Another proof of electron trajectories was obtained by XPS (X-ray Photoelectron Spectroscopy) test, F element was detected at the flashover traces on the face of the ceramic window. The element is not contained in the coupler and the window device itself but only existed on the internal surface of the cavity as residue from hydrofluoric acid after buffer chemistry polishing. The test results of the window control sample installed on the same cavity without field emission effect during the operation process show that there is no F element on the window. Because the F element escape from the cavity surface due to the starburst caused by electron emission [9], then it reaches the window surface under the action of the electric field. The FE electrons, which are also negative, will move along with the F element towards the window surface.

The secondary electron multiplication model is difficult to explain the flashover phenomenon described in this article. Firstly, it is generally accepted that field emission electrons at the cathode-insulator-vacuum junction insulator surface charging are involved in SEEA type surface flashover [5]. However, the coupling requirement of the coaxial pickup of the HWR type SRF cavity is extremely low ($\beta e=1/1000$), the power coupled to the window face is approximately 30mW($Pe/Pc=\beta e$, $Pc=\omega Uo/Qo$) [6] The electric potential between the 4mm distance of inner and outer conductors is about 15V (Simulation results of electromagnetic software for cavity field). It is not strong enough to excite field emission electrons from one side of window face to fly through. It is well known, for vacuum breakdown, SEEA type of flashover is usually observed in an electric field of at least MV/m [10]. Therefore, the weak voltage provided by a weak RF field in the PT coupler is unlikely to occur on the window.

Secondly, the ionization of the gas layer is also a way of SEEA flashover. In order to clean the environment inside the cavity and obtain good performance, the superconducting cavity has been baked at 120°C before going online to completely exhaust the gas adsorbed by the metal cavity wall. In the PT coupler, the $Al_2O_3$ ceramic window is a dense material, which is difficult to absorb and store gas. The vacuum in the superconducting cavity include PT coupler maintains of the order of $1\times10^{-7}$ Pa with little gas evolution during the entire operation. There almost no gas source to provide sufficient gas layer required for high-speed electron proliferation and diffusion in the prosecco of SEEA-type flashover.

Finally, the secondary electron yield (SEY) is also a key parameter of the secondary electron multiplication model. In this paper, the SEY coefficient of the ceramic window was tested. While the energy of incident electrons increases, the SEY coefficient tests a monotonically decreasing trend line when the maximum energy of incident electrons reaches 30000 eV, the SEY coefficient of the ceramic window was tested close to zero. The electrons reached the window surface with the energy above 0.3MeV have virtually impossible to give above-unity yields and establish the first traces of charge becoming the course of the entire flashover mechanism.

In contrast, the PER is consistent with all the observations. In this case, the flashover occurring without high voltage applied to the poles of the window and without the participation of a large amount of gas, and only relatively high-energy electron beam bombardment triggers the flashover. This phenomenon is consistent with the situation described by the PER model [11-13], the theory of the Polarization Energy Relaxation around trapped Electrons that are detrapped. After processing the experimental data, it is found that the maximum amount of charge discharged during flashover occurrence is no more than a few hundred nC. The flashover with such a small

amount of charge discharge can be explained by the PER model, which is an application of the polaron model

According to this model, electrons of the order of 0.3 MeV are trapped at the end of the penetration path, several microns below the surface and the breakdown propagate along the trapping sites. Specifically, the potential energy of electrons in the trap of high magnetic susceptibility area trapped all the neighboring atoms polarized and displaced. The weak repulsive force between trapped and polarons makes a high range space charge density stay at a temporary steady state before emergence of a perturbance, until the impact of another external electrons after. Once it is triggered, the temporary steady state is broken accompany with charge detrapping and relaxation of energy of polarization, which lead to electron avalanche process. The mechanism of this model describes the process relies on polarization participation instead of ionization, so there is few escape of electrons in the material itself and no large amount of free electron proliferation to participate in the propagation at the final breakdown. This is consistent with the experimental data that the amount of charge discharged by flashover is very small.

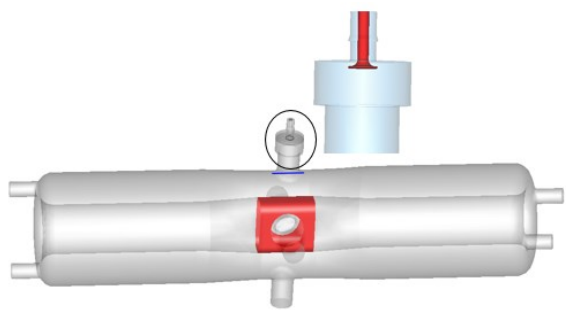

Figure5:The antenna with a big top is applied across the pick-up coupler.

According to the above analysis, field emission electrons from cavity fling through the space between the Ti material straight rod antenna and outer conductor of PT coupler excited flashover by hitting and accumulating on the face of window. Therefore, a direct and effective solution to the flashover is to avoid the accumulation of the FE electrons on the ceramic window. Making the antenna added with a diameter of 15mm to completely cover the Pt coupler port and prevent FE electrons from entering. The simulations on the newly designed antenna show that the electrons beams can be completely blocked by the antenna head without penetrated electrons and this structure designed will not trigger multipacting. Specifically, the escape rate of secondary electrons excited from the antenna by injected electrons is 4.79%, among which the rate of electrons can still reach the window is 4.43E-5%. The vertical test of the SRF cavity with the new designed antenna verified that the antenna does not cause multipacting and any other problems that affect the performance of the cavity. After the antenna was applied online, the previous similar flashing phenomenon was not detected during the operation of the accelerator for one year.

## ACKNOWLEDGMENT

This work was supported by The National Science Fund for Distinguished Young Scholars (Grant NO.Y836050WR0). The authors wish to thank Guirong Huang for the guidance. We also would like to acknowledge the helpful discussions with Changhua Chen and Xianchen Bai from the Science and Technology on High Power Microwave Laboratory, and the important consulting with Wanzhao Cui from China Academy of Space Technology.

# REFERENCES

[1] Shinian Fu, "Proton linac for ADS application in China" Proceedings of Linear Accelerator Conference LINAC2010, Tsukuba, Japan TUP019

[2] Yuan He, "Development of accelerator advanced nuclear energy (ADANES) and nuclear fuel recycle" IPAC2019, Melbourne, Australia

[3] R. A. Anderson, "Mechanism of fast surface flashover in vacuum*" Appl. Phys. Lett. 24, 54 (1974)

[4] Weiming Yue, "Development of a low beta half-wave superconducting cavity and its improvement from mechanical point of view" NIMA 163259 (2019);

[5] H. Craig Miller, "Flashover of Insulators in vacuum review of the phenomena and techniques to improve holdoff voltage" IEEE Trans. Electr. Insul. Vol.28 No. 4, August 1993

[6] H. Padamsee, J. Knobloch, T. Hays, "RF Superconductivity for Accelerators", JohnWiley & Sons, New York, 1998, pp.146-164

[7] R. H. Fowler, L. Nordheim, "Electron emission in intense electric fields", Proc. R. Soc. Lond. A, Math. Phys. Sci., 119:173(1928)

[8] T. L. Kirk, O. Scholder, "Evidence of nonplanar field emission via secondary electron detection in near field emission scanning electron microscopy" Appl. Phys. Lett. 94, 153502 （2009）

[9] H. Padamsee, J. Knobloch, "The nature of field emission from microparticles and the ensuing voltage breakdown", AIP Conference Proceedings 474, 212 (1999)

[10] H. Craig Miller, "Flashover of insulators in vacuum: the last twenty years", IEEE Transactions on Dielectrics Electrical Insulation Vol. 22 No. 6, December 2015

[11]G. Blaise, C. Le Gressus, "Charging and flashover induced by surface polarization relaxation process", Journal of Applied Physics 69, 6334 (1991) G. Blaise, IEEE Trans. Electron. Insul. **28**, 437 ~1993!

[12] H. Gong, C. K. Ong, and C. Le Gressus, "New observation of trapped charge transportation on circularly bound polymethylmethacrylate surface", Appl. Phys. Lett. 67, 2243 (1995)

[13] G. Blaise, C. Le Gressus, "Electron-trapping and energy localization in insulating materials", Received 9 July 2018